\documentclass[submission,copyright,creativecommons]{eptcs}
\providecommand{\event}{GASCom 2024} 

\usepackage{iftex}
 \usepackage[utf8]{inputenc}
 
\usepackage[english]{babel}
\usepackage{color}

\definecolor{red}{rgb}{1,0,0}
\definecolor{orange}{rgb}{1,0.5,0}

\usepackage[utf8]{inputenc}
\usepackage{amsmath}
\usepackage{amsthm}
\usepackage{enumerate}
\usepackage[warn]{textcomp}
\usepackage{graphicx}
\usepackage{caption}
\usepackage{tikz}
\usepackage{hyperref}
\usepackage[normalem]{ulem}

\newtheorem{theorem}{Theorem}

\newtheorem{remark}[theorem]{Remark}
\newtheorem{note}[theorem]{Note}

\ifpdf
  \usepackage{underscore}         

  \usepackage[T1]{fontenc}        
\else
  \usepackage{breakurl}           
\fi

\title{Local  Limit Theorems \\
for 
 $q$-Multinomial 
 and Multiple Heine
  Distributions}
\author{Malvina Vamvakari
\institute{Department of Informatics
and Telematics\\
Harokopio University of Athens
\\
Greece}
\email{mvamv@hua.gr}}
\def\titlerunning{Local Limit Theorems for $q$-Multinomial and Multiple Heine Distributions}
\def\authorrunning{M. Vamvakari}
\begin{document}
\maketitle

\begin{abstract}
In this work we establish 
local
limit theorems  for 
 $q$-multinomial
and multiple Heine 
distributions. Specifically, the pointwise convergence of the $q$-multinomial
distribution  of the first kind,  as well as for 
its discrete limit, the multiple Heine distribution,  to a  multivariate Stieltjes-Wigert type distribution, are provided.
\end{abstract}

\section{
Brief 
Introduction}
Recently, Vamvakari \cite{malvina} introduced 
multivariate discrete $q$-distributions.
Specifically, she  derived a multivariate absorption distribution as a conditional distribution of a Heine process at a finite sequence of $q$-points in a time interval which had been defined by Kyriakoussis and Vamvakari \cite{kyrvam}. Also, she deduced a multivariate $q$-hypergeometric distribution, as a conditional distribution of the multivariate absorption distribution.
 \newline 
Afterwards, Charalambides \cite{Charal3,Charal4}  introduced   in detail $q$-multinomial, negative $q$-multinomial, multivariate $q$--P\'olya and inverse $q$--P\'olya distributions and also examined their limiting discrete distributions.
Analytically, he considered 
 a stochastic model of a sequence of independent Bernoulli trials with chain-composite successes (or failures),
 where the odds of success of a certain kind at a trial is assumed to vary geometrically, with rate $q$, with the number of previous trials and introduced the $q$-multinomial and negative $q$-multinomial distributions of the first kind as well as their discrete limit, multivariate Heine distribution. Also, he considered   a stochastic model of a sequence of independent Bernoulli trials with chain-composite successes (or failures), where the probability of success of a certain kind at a trial varies geometrically, with rate $q$, with the number of previous successes and introduced the $q$-multinomial and negative $q$-multinomial distributions of the second kind kind as well as their discrete limit, multivariate Euler distribution.
\newline
Kyriakoussis and Vamvakari \cite{kyriakmvamv,kyriakmvamvconfl, kyrvampom}  studied   the  continuous 
limiting behaviour 
of the
univariate discrete $q$-distributions. Analytically,  they established
 the pointwise convergence  of the $q$-binomial and the negative $q$-binomial distributions of the first kind, as well as  of the Heine distribution, to a deformed Stieltjes Wigert continuous one.
Moreover, they  proved  the pointwise convergence of the
 $q$-binomial and the negative $q$-binomial distributions of the second kind, as well as of the Euler   distribution, 
to a deformed Gaussian one.
 \newline
Vamvakari \cite{malvina2} initiated  the study of continuous limiting  behaviour of multivariate discrete $q$-distributions
 inspired by the limiting behaviour of the univariate  ones. Specifically,
 she  studied the asymptotic behavior of the  univariate, bivariate and multivariate 
 absorption discrete $q$-distributions. The pointwise convergence of the univariate absorption distribution to a deformed Gaussian one and that of the bivariate  and multivariate absorptions to a bivariate and multivariate  deformed  Gaussian ones, have been provided. 
 \newline
  The aim of this work is to study  further the continuous limiting behaviour of multivariate discrete $q$-distributions. 
Local
limit 
theorems  for  the $q$-multinomial and
multiple Heine
 distributions are established.
 Specifically, the pointwise convergence of the $q$-multinomial and its discrete limit, the multiple Heine distribution, to a multivariate Stieltjes-Wigert type distribution are provided.
\section {Preliminaries}
 Charalambides \cite{Charal2} had studied in details the $q$-binomial distribution of the first kind and its discrete limit, the Heine Distribution
with probability functions (p.f.) given by
\begin{equation}
\label{kempbinom} f_X^B(x)={ n \choose x }_q q^{{x \choose
2}}{\theta}^x \prod_{j=1}^{n}(1+\theta
q^{j-1})^{-1},\quad x=0,1,\ldots,n,
\end{equation}
where $\theta>0$ and $0<q<1 $ 
and
\begin{equation}
\label{heine} f_X^H(x)=e_q(-\lambda )\frac{ q^{{x \choose
2}}\lambda^x }{[x]_q!}, \quad x=0,1,2,\ldots, \ 0<q<1,\ 0<\lambda<\infty,
\end{equation}
where
\begin{equation}
\label{qexp}
e_q(z):=\sum_{n=0}^\infty \frac{(1-q)^n z^n}{(q;q)_n}=\sum_{n=0}^\infty \frac{ z^n}{[n]_q!}=\frac{1}{((1-q)z;q)_\infty},\quad |z|<1,
\end{equation}
and
\begin{equation}
\label{factorial}
[n]_{q}! = [1]_{q}[2]_{q}\cdots[n]_{q}= \prod_{k=1}^n\frac{1-q^k}{(1-q)^n}= \frac{(q;q)_n}{(1-q)^n},\quad 0<q<1, \quad [t]_{q}=\frac{1-q^{t}}{1-q}.
\end{equation}
 Kyriakoussis and Vamvakari \cite{kyriakmvamv, kyriakmvamvconfl} proved limit
theorems among others  for the $q$-Binomial distribution of the first kind
and Heine distribution
for constant~$q$, 
by using pointwise convergence in a ``$q$-analogous sense'' of
the classical de~Moivre--Laplace limit theorem. Specifically 
for
the needs of their study they established a $q$-Stirling formula for $n \rightarrow
\infty$ of the $q$-factorial of order $n$,  defined  by relation (\ref{factorial}).
Analytically,  for $0<q<1$ constant, it was proved that,  
\begin{equation}
\label{asymptexpanF} [n]_q! = \frac{q^{-1/8}(2 \pi(1-q))^{1/2}}{(q\log
q^{-1})^{1/2}}\frac{ q^{{n \choose
2}}q^{-n/2}[n]_{1/q}^{n+1/2}}{\prod_{j=1}^{\infty}(1+(q^{-n}-1)q^{j-1})}
 \left( 1 +O(n^{-1}) \right).
\end{equation}
\newline
Next, the
pointwise convergence of the $q$-Binomial distribution of the first kind to a
deformed continuous Stieltjes--Wigert distribution was established. 
The continuous Stieltjes--Wigert distribution has
probability density function
\begin{equation}
\label{wigert} v_W^{SW}(w)=\frac{q^{1/8}}{\sqrt{2 \pi \log
q^{-1}\,w}} e^ \frac{ (\log w)^2}{2 \log q},\quad w>0,
\end{equation}
with mean value $\mu^{SW}=q^{-1}$ and standard deviation
$\sigma^{SW}=q^{-3/2}(1-q)^{1/2}$.
\newline
Transferred from the random variable $X$ of the
$q$-Binomial distribution \eqref{kempbinom} to the
equal-distributed deformed random variable $Y=[X]_{1/q}$ and
for $n\rightarrow \infty$, the $q$-Binomial distribution of the first kind was
approximated by a deformed standardized continuous
Stieltjes--Wigert distribution as follows:
\begin{align}
f_X^B(x) &\cong \frac{q^{-7/8}}{\sigma_q (2 \pi)^{1/2}}\left(\frac{\log q^{-1}}{q^{-1}-1}\right)^{1/2}\left(q^{-3/2}(1-q)^{1/2}\frac{[x]_{1/q}-\mu_q}{\sigma_q}+q^{-1}\right)^{-1/2}q^{-x}\nonumber\\
&\cdot
\exp\left(\frac{1}{2\log q}\log^2 \left(
 q^{-3/2}(1-q)^{1/2}\frac{[x]_{1/q}-\mu_q}{\sigma_q}+q^{-1}
 \right)\right),\quad x\geq 0,
\label{approxheineFin}
\end{align}
where $\theta=\theta_n$, for $n=0,1,2,\ldots$, such that
$\theta_n=q^{-\alpha n}$ with $0<a<1$ constant and $\mu_q$ and
$\sigma_q^2 $ the mean value and variance of the random variable
$Y,$ respectively. A similar asymptotic result has been provided 
 for the Heine distribution when $\lambda \rightarrow \infty.$

\section{Main Results}
Let $X_{j}$ be the number of successes of a $j$th  kind in a sequence of $n$ independent Bernoulli trials with chain composite failures, where the 
probability of success of the $j$th kind at the $i$th trial
 is given by
\[p_{j,i}=\frac{\theta_jq^{i-1}}{1+\theta_jq^{i-1}}, \,0<\theta_j<\infty,\,\,j=1,2,\ldots, i=1,2,\ldots,\,0<q<1 \,\,\,\mbox{or}\, \,\,1<q<\infty.\]
\newline
Then the joint probability function of the random vector ${\cal X}=\left(X_{1},X_{2},\ldots,X_{k}\right)$  is given by
\begin{eqnarray}
\label{qmultinom}
f^B_{{\cal X}}(x_1,x_2,\ldots,x_k)=
P(X_{1}=x_1,X_{2}=x_2,\ldots,X_{k}=x_k)
=
{ n \choose {x_1,x_2,\ldots,x_k}}_q 
\prod_{j=1}^k\frac{\theta_j^{x_j}q^{x_j \choose 2}}{\prod_{i=1}^{n-s_{j-1}}(1+\theta_jq^{i-1})}
\end{eqnarray}
$x_j=0,1,2,\dots,n$, 
$\sum_{j=1}^k x_j\leq n$, $s_j=\sum_{i=1}^jx_i,$ $\,\, 0<\theta_j<1,\, j=1,2,\ldots,k,\,\,\mbox{and}\,\, 0<q<1\,\,\mbox{or}\,\,1<q<\infty.$
This discrete $q$-distribution is known as a {\em $q$-multinomial distribution}  (see Charalambides \cite{Charal3}).
\newline
The discrete  limit of the joint  p.f. of the $q$-multinomial distribution of the $1$st kind, as $n \rightarrow \infty$, is the joint p.f. of the {\em multiple Heine distribution},
\begin{eqnarray}
\label{multheine}
\lim_{n \rightarrow \infty}{ n \choose {x_1,x_2,\ldots,x_k}}_q 
\prod_{j=1}^k\frac{\theta_j^{x_j}q^{x_j \choose 2}}{\prod_{i=1}^{n-s_{j-1}}(1+\theta_jq^{i-1})}=\prod_{j=1}^k\frac{q^{{x_j \choose
2}}{\lambda_j}^x }{[x_j]_q!}\prod_{i=1}^{\infty}(1+\lambda_j (1-q)
q^{i-1})^{-1}
\end{eqnarray}
$x_j=0,1,2,\dots$, $\,\lambda_j>0,\,0<q<1$, $\,\lambda_j=\theta_j/(1-q)$, $\,j=1,2,\ldots,k$  (see Charalambides \cite{Charal3}).
 \newline
Next we will study the continuous limiting behaviour of the {\it $q$-trinomial} distribution.
Let $\left(X_{1},X_{2} \right)$ be the discrete bivariate  random variable  with joint probability function
\begin{eqnarray}
\label{qtrivar}
f^B_{X_{1},X_{2}}(x_1,x_2)=P\left(X_{1},X_{2} \right)={n \choose {x_1,x_2}}_q\,\,
\frac{\theta_1^{x_1}\theta_2^{x_2}q^{{x_1 \choose 2}+{x_2 \choose 2}}}{\prod_{i=1}^n(1+\theta_1q^{i-1})\prod_{i=1}^{n-x_1}(1+\theta_2q^{i-1})}
\end{eqnarray}
 $x_j=0,1,2,\ldots,n$, $j=1,2,$ $x_1+x_2\leq n$,$\,\, 0<\theta_1, \theta_2<1\,\,\mbox{and}\,\, 0<q<1\,\,\mbox{or}\,\,1<q<\infty.$
 The distribution of the   bivariate  random variable $\left(X_{1},X_{2}\right)$  is known as a {\it $q$-trinomial distribution}.
 \newline
 The marginal probability function of the random variable $X_1,$ is distributed according to the  $q$-binomial of the $1$st kind with probability function
\begin{eqnarray}
\label{bivarabs2b}
f^B_{X_{1}}(x_1)={n \choose {x_1}}_q \frac{\theta_1^{x_1}q^{x_1 \choose 2}}{\prod_{i=1}^n(1+\theta_1q^{i-1})},
\,\,\ x_1=0,1,2,\ldots,n.\nonumber
\end{eqnarray}
\newline
The mean and the variance of the deformed variable $[X_{1}]_{1/q}$ are given by
\begin{eqnarray}
\label{meanvarYEb2}
 \mu_{[X_{1}]_{1/q}}&=&E\left([X_{1}]_{1/q}\right)=
 [n]_{q}\frac{\theta_1}{1+\theta_1
q^{n-1}} \nonumber\\
\quad &&\mbox{and} \quad \\
 (\sigma_{[X_{1}]_{1/q}})^2&=&V\left([X_{1}]_{1/q}\right)=\frac{1-q}{q}[n]_q^2\frac{\theta_1^2}{(1+\theta_1
q^{n-1})^2(1+\theta_1 q^{n-2})}+[n]_q\frac{\theta_1}{(1+\theta_1
q^{n-1})(1+\theta_1 q^{n-2})}
 \nonumber
\end{eqnarray}
respectively.
\newline
The conditional random variable $X_{2}|X_{1},$ is distributed according to the univariate 
  $q$-binomial of the $1$st kind
with probability function
\begin{eqnarray}
\label{bivarabs2a}
f^B_{X_{2}|X_{1}}(x_2|x_1)={n-x_1 \choose {x_2}}_q \frac{\theta_2^{x_2}q^{x_2 \choose 2}}{\prod_{i=1}^{n-x_1}(1+\theta_2q^{i-1})},
\,\,\ x_2=0,1,2,\ldots,n-x_1.\nonumber
\end{eqnarray}
\newline
The conditional mean and conditional variance  of the deformed variable $[X_{2}]_{1/q}$ given $X_{1}=x_1$, are given by
\begin{eqnarray}
\label{meanvarYEa2}
 \mu_{[X_{2}]_{1/q}|X_{1}}&=&
  E\left([X_{2}]_{1/q}|X_1\right)= [n-x_1]_{q}\frac{\theta_2}{1+\theta_2
q^{n-x_1-1}},\,\,
\nonumber\\
 \quad &&
 \mbox{and} \quad \\
(\sigma_{[X_{2}]_{1/q}|X_{1}})^2&=&V\left([X_{2}]_{1/q}|X_1\right)=
\frac{1-q}{q}[n-x_1]_q^2\frac{\theta_2^2}{(1+\theta_2
q^{n-x_1-1})^2(1+\theta_2 q^{n-x_1-2})}\nonumber\\&&+[n-x_1]_q\frac{\theta_2}{(1+\theta_2
q^{n-x_1-1})(1+\theta_2 q^{n-x_1-2})},
 \nonumber
\end{eqnarray}
respectively.
\begin{note}
The conditional $q$-mean,  $\mu_{[X_{2}]_{1/q}|X_{1}}$, can be interpreted as a  \emph{$q$-regression curve}.
\end{note}
Let us now consider the deformed  random variables
$[X_{1}]_{1/q}$ and $[X_{2}]_{1/q}$  as well as the
$q$-standardized random variables
 $Z=\frac{[X_{1}]_{1/q}-\mu_{[X_{1}]_{1/q}}}{\sigma_{[X_{1}]_q}}$
 and
  $W=\frac{[X_{2}]_{1/q}- \mu_{[X_{2}]_{1/q}|X_{1}}}{\sigma_{[X_{2}]_{1/q}|X_{1}}}$ with
$\mu_{[X_{1}]_{1/q}}$,$\sigma_{[X_{1}]_{1/q}}$ and
$\mu_{[X_{2}]_{1/q}|X_{1}}$, $\sigma_{[X_{2}]_{1/q}|X_{1}}$
 given by \eqref{meanvarYEb2}
 and \eqref{meanvarYEa2}, respectively.   
 Then,   we apply pointwise convergence techniques to the joint probability function (\ref{qtrivar}), by  using suitably the $q$-Stirling type
  (\ref{asymptexpanF}),
 and we obtain the following theorem concerning the asymptotic behaviour of the 
$q$-trinomial distribution.
\newline
\begin{theorem} \label{Vthm:1} 
Let $\theta_1=\theta_{1,n}=q^{-\alpha_1 n}$ and $\theta_2=\theta_{2,n}=q^{-\alpha_2 n}$ with  $0<a_1,\, a_2<1$ constants and $0<q<1$.
 Then, for $n \rightarrow \infty$, the $q$-trinomial distribution of the first kind
 is approximated 
 by  a deformed standardized bivariate continuous
Stieltjes-Wigert distribution 
 as follows:
 {\small
\begin{eqnarray}
\label{approxqbinBS2}f^B_{X_{1},X_{2}}(x_1,x_2) &\cong&
\frac{q^{-7/4} {\log
q^{-1}}}{2 \pi(q^{-1}-1)\sigma_{[X_{1}]_{1/q}}\sigma_{[X_{2}]_{1/q}|X_{1}}}
q^{-(x_1+x_2)}\left(q^{-3/2}(1-q)^{1/2}\frac{[x_1]_{1/q}-\mu_{[X_{1}]_{1/q}}}{\sigma_{[X_{1}]_{1/q}}}+q^{-1}\right)^{-1/2}\nonumber\\
&&\cdot\left(q^{-3/2}(1-q)^{1/2}\frac{[x_2]_q- \mu_{[X_{2}]_{1/q}|X_{1}}}{\sigma_{[X_{2}]_{1/q}|X_{1}}}+q^{-1}\right)^{-1/2}\nonumber\\
&&\cdot \exp\left(\frac{1}{2\log q}\left(\log^2 \left(
 q^{-3/2}(1-q)^{1/2}\frac{[x_1]_{1/q}-\mu_{[X_{1}]_{1/q}}}{\sigma_{[X_{1}]_{1/q}}}+q^{-1}
 \right)
 \right)\right)
 \nonumber\\
 &&\cdot  \exp\left(\frac{1}{2\log q}\log^2 \left(
 q^{-3/2}(1-q)^{1/2}\frac{[x_2]_q- \mu_{[X_{2}]_{1/q}|X_{1}}}{\sigma_{[X_{2}]_{1/q}|X_{1}}}+q^{-1}
 \right)\right), \, \, x_1,\,x_2\geq 0,
\end{eqnarray}}
where
$\mu_{[X_{1}]_{1/q}}$ and $\sigma^2_{[X_{1}]_{1/q}},$ 
  given in \eqref{meanvarYEb2}, are the mean value and the variance of the random variable $[X_1]_{1/q}$ while
  $\mu_{[X_{2}]_{1/q}|X_{1}}$ and $\sigma^2_{[X_{2}]_{1/q}|X_{1}},$
    given in 
\eqref{meanvarYEa2}, are the conditional  mean value and the conditional  variance of the random variable
$[X_{2}]_{1/q}$ given $X_{1}=x_1$.
\end{theorem}
Next we expand our study on the asymptotic behaviour of the $q$-multinomial distribution with  joint p.f. (\ref{qmultinom}).
\newline
 The marginal probability function of the random variable $X_1,$ is distributed according to the  $q$-binomial of the $1$st kind with probability function
\begin{eqnarray}
\label{bivarabs2bc}
f^B_{X_{1}}(x_1)={n \choose {x_1}}_q \frac{\theta_1^{x_1}q^{x_1 \choose 2}}{\prod_{i=1}^n(1+\theta_1q^{i-1})},
\,\,\ x_1=0,1,2,\ldots,n.\nonumber
\end{eqnarray}
\newline
The mean and the variance of the deformed variable $[X_{1}]_{1/q}$ are given by
\begin{eqnarray}
\label{meanvarYEb2a}
 \mu_{[X_{1}]_{1/q}}&=&E\left([X_{1}]_{1/q}\right)=
 [n]_{q}\frac{\theta_1}{1+\theta_1
q^{n-1}} \nonumber\\
\quad &&\mbox{and} \quad \\
 (\sigma_{[X_{1}]_{1/q}})^2&=&V\left([X_{1}]_{1/q}\right)=\frac{1-q}{q}[n]_q^2\frac{\theta_1^2}{(1+\theta_1
q^{n-1})^2(1+\theta_1 q^{n-2})}+[n]_q\frac{\theta_1}{(1+\theta_1
q^{n-1})(1+\theta_1 q^{n-2})}
 \nonumber
\end{eqnarray}
respectively.
\newline
The conditional random variables  $X_{2}|X_1, X_{3}|(X_{1},X_2), \ldots, X_k|(X_1,\ldots,X_{k-1})$ are distributed according to univariate $q$-binomial distributions of the $1$st kind
with probability functions
\begin{eqnarray}
\label{bivarabs2am}
f^B_{X_k|(X_1,\ldots,X_{k-1})}(x_k|x_1,x_2,\ldots,x_{k-1})&=&{n-\sum_{j=1}^{k-1}x_j \choose {x_k}}_q \,\,\frac{\theta_k^{x_k}q^{x_k \choose 2}}{\prod_{i=1}^{n-\sum_{j=1}^{k-1}x_j}(1+\theta_kq^{i-1})},\nonumber\\&&\quad x_k=0,1,\ldots,\sum_{j=1}^{k-1}x_j,\,\,k \geq 2.\nonumber
\end{eqnarray}
The conditional mean and conditional variance  of the deformed variables $[X_{j}]_{1/q}$ given $X_{1}=x_1,\ldots,$ $X_{j-1}=x_{j-1}$, $j=2,\ldots,k,\,\,$ $k\geq 2,$ are given respectively  by
\begin{eqnarray}
\label{meanvarYEam}
 \mu_{[X_{j}]_{1/q}|(X_{1},\ldots,X_{j-1})}&=&
  E\left([X_{j}]_{1/q}|(X_{1},\ldots,X_{j-1})\right)= [n-s_{j-1}]_{q}\frac{\theta_j}{1+\theta_j
q^{n-s_{j-1}-1}}\,\,
\nonumber\\
\mbox{and} \quad \\
\sigma^2_{[X_{j}]_{1/q}|(X_{1},\ldots,X_{j-1})}&=&V\left([[X_{j}]_{1/q}|(X_{1},\ldots,X_{j-1})\right)\nonumber\\&=&
\frac{1-q}{q}[n-s_{j-1}]_q^2\frac{\theta_j^2}{(1+\theta_j
q^{n-s_{j-1}-1})^2(1+\theta_j q^{n-s_{j-1}-2})}\nonumber\\
&&+[n-s_{j-1}]_q\frac{\theta_j}{(1+\theta_j
q^{n-s_{j-1}-1})(1+\theta_j q^{n-s_{j-1}-2})},
 \nonumber
\end{eqnarray}
where $s_{j-1}=\sum_{i=1}^{j-1}x_i$, $\,\,j=2,\ldots,k$, $k\geq 2$.
\begin{note}
 It should be noted that the conditional $q$-means,  $\mu_{[X_j]_{1/q}|(X_1,\ldots,X_{j-1})}$, $3 \leq j \leq k,$ $k\geq 3,$ can be interpreted as  $q$-regression hyperplanes.
 \end{note}
Let us now consider the deformed  random variables
\[[X_{j}]_{1/q},\,j=1,\ldots,k,\,\,k \geq 1\]
and the
$q$-standardized random variables
 \[Z_1=\frac{[X_{1}]_{1/q}-\mu_{[X_{1}]_{1/q}}}{\sigma_{[X_{1}]_q}},\,\,
  Z_j=\frac{[X_{j}]_{1/q}-  \mu_{[X_{j}]_{1/q}|(X_{1},\ldots,X_{j-1})}}{\sigma_{[X_{j}]_{1/q}|(X_{1},\ldots,X_{j-1})}}, \,j=2,\ldots,k, \,\, k \geq 3,\]  with
$\mu_{[X_{1}]_{1/q}}$,$\sigma_{[X_{1}]_{1/q}}$ and
$\mu_{[X_{j}]_{1/q}|(X_{1},\ldots,X_{j-1})}$,  $\sigma_{[X_{j}]_{1/q}|(X_{1},\ldots,X_{j-1})}$
 given by \eqref{meanvarYEb2a}
 and \eqref{meanvarYEam}, respectively. Then,   we apply pointwise convergence techniques to the joint probability function (\ref{qmultinom}), by  using suitably the $q$-Stirling type
  (\ref{asymptexpanF}),
 and we obtain the following theorem concerning the asymptotic behaviour of the 
$q$-multinomial distribution.
\begin{theorem}
Let $\theta_j=\theta_{j,n}=q^{-\alpha_j n}$ 
 with  $0<a_j<1,$ $j=1,2,\ldots,k$ constants and $0<q<1$.
 Then, for $n \rightarrow \infty$, the $q$-multinomial distribution
 is approximated by a deformed multivariate standardized
   continuous
Stieltjes-Wigert distribution
distribution as follows:
{\small
\begin{eqnarray}
\label{approxqmultinom}&&f^B_{{\cal X}}(x_1,x_2,\ldots,x_k) \cong \left(\frac{q^{-7/8} {(\log
q^{-1})}^{1/2}}{(2 \pi)^{1/2}(q^{-1}-1)^{1/2}}\right)^k
\frac{ q^{-\sum_{j=1}^k x_j}}{\sigma_{[X_{1}]_{1/q}}\prod_{j=2}^k{ \sigma_{[X_{j}]_{1/q}|(X_{1},\ldots,X_{j-1})}}}\nonumber\\
&&
\quad\cdot\left(q^{-3/2}(1-q)^{1/2}\frac{[x_1]_{1/q}-\mu_{[X_{1}]_{1/q}}}{\sigma_{[X_{1}]_{1/q}}}+q^{-1}\right)^{-1/2}\nonumber\\
&&\quad\cdot\prod_{j=2}^k\left(q^{-3/2}(1-q)^{1/2}\frac{[x_{j}]_{1/q}- \mu_{[X_{j}]_{1/q}|(X_{1},\ldots,X_{j-1})}}{\sigma_{[X_{j}]_{1/q}|(X_{1},\ldots,X_{j-1})}}+q^{-1}\right)^{-1/2}\nonumber\\
&&
\quad\cdot
\exp\left(\frac{1}{2\log q}\left(\log^2 \left(
\frac{(1-q)^{1/2}}{q^{3/2}}\frac{[x_1]_{1/q}-\mu_{[X_{1}]_{1/q}}}{\sigma_{[X_{1}]_{1/q}}}+q^{-1}
 \right)
 \right)\right)
 \nonumber\\&&
 \quad\cdot\exp\left(\frac{1}{2\log q}\sum_{j=2}^k\log^2 \left(
\frac{(1-q)^{1/2}}{q^{3/2}}\frac{[x_{j}]_{1/q}- \mu_{[X_{j}]_{1/q}|(X_{1},\ldots,X_{j-1})}}{\sigma_{[X_{j}]_{1/q}|(X_{1},\ldots,X_{j-1})}}+q^{-1}
 \right)\right),
 \nonumber\\
&&\quad\quad\quad\ x_j \geq 0, j=1,2,\ldots,k, k\geq 2, 
\end{eqnarray}
}
where
$\mu_{[X_{1}]_{1/q}}$ and $\sigma^2_{[X_{1}]_{1/q}},$ 
  given in \eqref{meanvarYEb2a}, are the mean value and the variance of the random variable $[X_1]_{1/q},$ while
$ \mu_{[X_{j}]_{1/q}|(X_{1},\ldots,X_{j-1})}$  and $\sigma^2_{[X_{j}]_{1/q}|(X_{1},\ldots,X_{j-1})},$
    given in 
\eqref{meanvarYEam}, are the conditional  mean values and the conditional  variances of the random variables
$[X_{j}]_{1/q}$ given $X_{1}=x_1,\ldots,X_{j-1}=x_{j-1}, $ $\,j=2,\ldots,k$, $\,k\geq 2$.
\end{theorem}
\begin{remark} Let  ${\cal X}=\left(X_{1},X_{2},\ldots,X_{k}\right)$ be a random vector that follows the multiple Heine distribution,  defined in (\ref{multheine}). Then the  joint p.f. the multiple Heine distribution is  given by
\begin{eqnarray}
\label{multheine2}
f^H_{{\cal X}}(x_1,x_2,\ldots,x_k)
=\prod_{j=1}^k\frac{q^{{x_j \choose
2}}{\lambda_j}^x }{[x_j]_q!}\prod_{i=1}^{\infty}(1+\lambda_j (1-q)
q^{i-1})^{-1}, \nonumber
\end{eqnarray}
where $x_j=0,1,2,\dots$, $\,\lambda_j>0,\,0<q<1$, $\,\lambda_j=\theta_j/(1-q)$, $\,j=1,2,\ldots,k,\,\,k\geq 2.\,\,$  
\newline
Since the random variables $X_j$, $\,j=1,2,\ldots,k$, $\,k \geq 2,$ are independent, we easily  derive that, 
for $\lambda_j \rightarrow \infty$, $\,j=1,2,\ldots,k$, the multiple Heine distribution
 is approximated by a deformed multivariate standardized
   continuous
Stieltjes-Wigert distribution
distribution as follows:
{\small
\begin{eqnarray}
\label{approxqmultinomb}f^H_{{\cal X}}(x_1,x_2,\ldots,x_k)& \cong& \left(\frac{q^{-7/8} {(\log
q^{-1})}^{1/2}}{(2 \pi)^{1/2}(q^{-1}-1)^{1/2}}\right)^k
\frac{ q^{-\sum_{j=1}^k x_j}}{\prod_{j=1}^k\sigma_{[X_{j}]_{1/q}}
}
\nonumber\\
&&\cdot
\prod_{j=1}^k\left(q^{-3/2}(1-q)^{1/2}\frac{[x_j]_{1/q}-\mu_{[X_{j}]_{1/q}}}{\sigma_{[X_{j}]_{1/q}}}+q^{-1}\right)^{-1/2}
\nonumber\\
&&
\cdot
\exp\left(\frac{1}{2\log q}\left(\sum_{j=1}^k\log^2
\frac{(1-q)^{1/2}}{q^{3/2}}\frac{[x_j]_{1/q}-\mu_{[X_{j}]_{1/q}}}{\sigma_{[X_{j}]_{1/q}}}
 \right)\right),\nonumber\\
 &&\quad\quad
 x_j \geq 0, j=1,\ldots,k, k\geq 2,
\end{eqnarray}
}
where
$\mu_{[X_{j}]_{1/q}}=\lambda_j$ and $\sigma^2_{[X_{j}]_{1/q}}=\lambda_jq^{-1}(1-q)+\lambda_j,$ $j=1,2,\ldots, k,$
   are respectively  the mean values and the variances of the random variables $[X_j]_{1/q},$ $\,j=1,2,\ldots,k.$

\end{remark}

\nocite{*}
\bibliographystyle{eptcs}
\bibliography{Vamvakari2}

\end{document}